\def\year{2021}\relax
\documentclass[letterpaper]{article} 

\usepackage{ulem}  
\usepackage{soul}   
\usepackage{xcolor} 
\usepackage{subfigure}
\usepackage[linesnumbered,ruled,vlined]{algorithm2e}
\usepackage{booktabs}

\usepackage{aaai21}  
\usepackage{times}  
\usepackage{helvet} 
\usepackage{courier}  
\usepackage[hyphens]{url}  
\usepackage{graphicx} 
\usepackage{natbib}  
\usepackage{caption} 
\title{KAST: Knowledge Aware Adaptive Session Multi-Topic Network for Click-Through Rate Prediction}

\author {
    Dike Sun\textsuperscript{\rm 1}, Kai Liu\textsuperscript{\rm 2}, ShengKai Yang\textsuperscript{\rm 2}\\
}

\affiliations{
    \textsuperscript{\rm 1} University of Electronic Science and Technology of China, Chengdu, China \\
    \textsuperscript{\rm 2} Alibaba Group, Hangzhou, China \\
    sundike@std.uestc.edu.cn, baiyang.lk@alibaba-inc.com, shengkai.ysk@alibaba-inc.com \\
}

\begin{document}
\maketitle
\frenchspacing


\begin{abstract}
Capturing the evolving trends of user interest is important for both recommendation systems and advertising systems, and user behavior sequences have been successfully used in Click-Through-Rate(CTR) prediction problems. However, if the user interest is learned on the basis of item-level behaviors, the performance may be affected by the following two issues. Firstly, some casual outliers might be included in the behavior sequences as user behaviors are likely to be diverse. Secondly, the span of time intervals between user behaviors is random and irregular, for which a RNN-based module employed from NLP is not perfectly adaptive. To handle these two issues, we propose the Knowledge aware Adaptive Session multi-Topic network(KAST). It can adaptively segment user sessions from the whole user behavior sequence, and maintain similar intents in the same session. Furthermore, in order to improve the quality of session segmentation and representation, a knowledge-aware module is introduced so that the structural information from the user-item interaction can be extracted in an end-to-end manner, and a marginal based loss with these information is merged into the major loss. Through extensive experiments on public benchmarks, we demonstrate that KAST can achieve superior performance than state-of-the-art methods for CTR prediction, and key modules and hyper-parameters are also evaluated.
\end{abstract}

\section{Introduction}

Many deep-learning-based methods have been proposed to model user behaviors and predict click-through rate(CTR) in the raking stage of recommendation systems and advertising systems, which obtain better online results than classic models, such as Logistic Regression(LR)\cite{kleinbaum2002logistic}, Factorization Machines(FM)\cite{rendle2010factorization}, etc. Wide\&Deep\cite{cheng2016wide}, DeepFM\cite{guo2017deepfm}, Deep\&Cross Net(DCN)\cite{wang2017deep} focuses on the intersection of category features, and produced outstanding results on large-scale sparse datasets in industry. In e-commerce systems, users generate a large amount of behavior data everyday, which can be utilized to greatly enhance the prediction effect. Models such as DIN\cite{zhou2018deep} and DIEN\cite{zhou2019deep} learn the user interest by integrating Natural Language Processing(NLP) techniques, such as Attention Mechanism\cite{bahdanau2014neural}, LSTM\cite{hochreiter1997long}, GRU\cite{chung2014empirical}, etc. 
However, some casual outliers might be included in the behavior sequences as user behaviors are likely to be diverse which reduces the effect of user interest extraction.

\begin{figure}[t]
  \setlength{\abovecaptionskip}{0pt} 
  \setlength{\belowcaptionskip}{-10pt}
  \centering
  \subfigure[Divide time gap: 10mins]{\includegraphics[width=4cm]{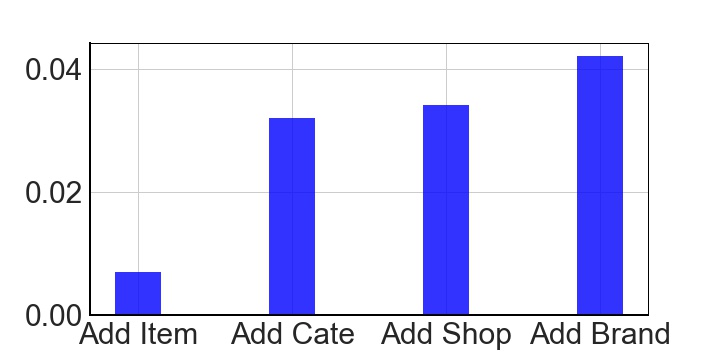}}
  \subfigure[Divide time gap: 30mins]{\includegraphics[width=4cm]{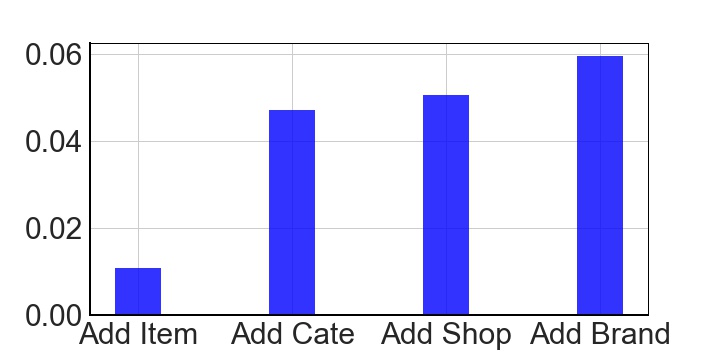}}
  \caption{Percentage of incorrectly divided sessions, where \emph{Item} means only \emph{Item} is used to judge the division, and \emph{Cate} means that is incorporated with \emph{Item}. Others can be similar explained.}
  \label{fig:pathdemo1}
\end{figure}

User behavior sequence consists of a series of sessions, and DSIN\cite{feng2019deep} pays attention into session-wise topic representation, rather than fine-grained item-wise representation. Meanwhile, the interest expressed within one session usually focuses on one topic. For instance, when a user wants to purchase a T-shirt, the topic of intent around this time window may be relevant to sport gears or cloths. In order to acquire high-quality interest topics, it is important to guarantee high-quality session division. At present, the common practice is that whenever a time gap of more than 30 minutes emerges, a division would be made\cite{grbovic2018real}. Although this method is simple and efficient, some items on the borders of adjacent sessions are likely to be mis-divided. Figure \ref{fig:pathdemo1} depicts the percentage of sessions that are incorrectly divided in Alibaba datasets \footnote{\url{http://www.cikm2019.net/challenge.html}} for CIKM2019 AnalytiCup, where the adjacent items have the same values for features (category), (category, shop), and (category, shop, brand). The time interval is set to 10 minutes and 30 minutes. It is shown that more than 4.2\%(6\%) of the items are misclassified by this method. Moreover, even if two items do not belong to the same category, shop or brand, they may still fall into the same topic, which means even more products are actually misclassified. Additionally, it is problematic to determine the interval between sessions, because both explicit and implicit similarities are crucial to the problem, and the latter ones are hard to be fully recognized.

In this paper, we propose a method called Knowledge aware Adaptive Session Multi-Topic Network(KAST), which employs two key modules: Adaptive Session Segmentation (ASS) module, and Knowledge aware Structure-information Extraction (KSE) module.  In the ASS module, on the basis of the original session, by adaptively segmenting the user behavior sequence into sessions, the effect of learning the session-level topic evolution can be enhanced, and interference from casual behaviors among the sequence is precluded, which the whole step is end-to-end. Moreover, the performance of ASS module depends on the quality of the embedding matrix. Therefore, it is expected that the more similar an item pair is in reality, the closer it will be in the latent space. Hence, we employ KSE module, which incorporates into KAST to optimize the user and item embedding matrices. The structural information between the users and items on the graph, as novel knowledge for embedding representation learning, is utilized in this process. In addition, the marginal based loss is merged into the major loss function. In this way, the KSE module further assists ASS module to accurately learn the topics of interest in session level. 
 
It should be noted that user behaviors in the sequence, such as click or collection, are not equally spaced. However, the input sequence in RNN is supposed to be equally spaced by default. Hence, having acquired the optimized sessions, the pooling layer distills the session behaviors into session-level vector representations. In this way, the non-equality problem between items within each session no longer jeopardizes the NLP conditions, and it is alleviated in the session-level. Next, GRU is utilized to capture the evolution of the topics, after which the attention layer is used to weigh all the sessions by evaluating the correlation between the target item and each session. In the end, the final representation is obtained.

To sum up, the main contributions of this paper are listed as follows:
\begin{itemize}
    \item We propose the KAST network architecture, which captures the topic of interests from each adaptively-divided session, and alleviates the problem 
    of unequal spacing in user behavior sequences. 
    \item We design an ASS module to update sessions automatically, which uses the dynamic updating embedding matrix to make the session division more reasonable, and reduces more manual feature engineering as well.
    \item In order to enhance the reliability of the ASS module, we further design the KSE module. It is able to learn the structural knowledge from the graph to improve the effect of the embedding matrix.
    \item We conduct extensive experiments to compare the proposed method with many typical methods on the public datasets, and we also evaluate the effect and robustness for ASS and KSE modules. It is shown that the proposed method obtains the state-of-art results on the CTR prediction task.

\end{itemize}
\begin{figure*}[ht]
\setlength{\abovecaptionskip}{10pt} 
\setlength{\belowcaptionskip}{-8pt}
\centering
\includegraphics[width=0.9\textwidth]{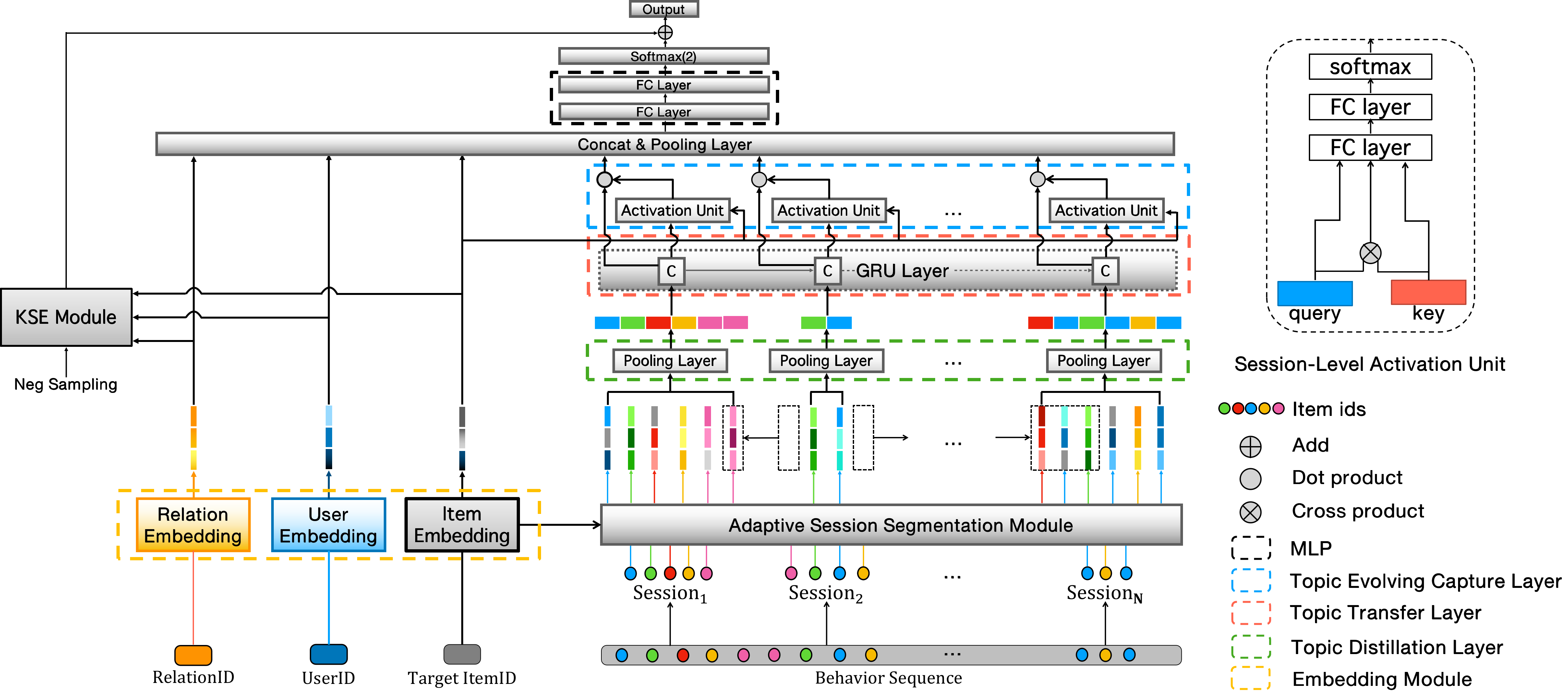}
\caption{The architecture of KAST}
\label{fig:pathdemo2}
\end{figure*}

\section{Related Work}

\subsection{General Deep Models}
Most deep networks are based on embedding and multi-layer perceptron(Emb\&MLP) structure, and Wide\&Deep\cite{cheng2016wide} combines the memory ability of the linear part and the generalization ability of the DNN part to improve the overall performance, which constructs the basis for most of subsequent deep models. Deep\&Cross Net(DCN)\cite{wang2017deep} can explicitly selects feature set to design higher-order feature crossing, which avoids useless combined features. The "cross" net structure can effectively learn the bounded-degree combined feature. Compared to Emb\&MLP, PNN\cite{qu2016product} designs the ``product layer'' after the embedding to capture field-based second-order feature correlation. AFM\cite{xiao2017attentional} adds attention mechanism on the basis of FM, which evaluates the importance of feature interactions and reduce the impact of feature noise. Similar to the Wide\&Deep, DeepFM\cite{guo2017deepfm} is also jointly trained by the shallow part and the deep part. The major difference is that the LR is replaced by FM in the shallow part, and FM is able to automatically learn cross features. AutoInt\cite{song2019autoint} uses multi-head self-attention mechanism to perform automatic feature-crossing learning, and therefore improves the accuracy of CTR prediction tasks.


\subsection{Sequence-based Deep Models}
The user's behavior sequence contains rich information, which implies user's interest trend. Therefore, modeling behavior sequences can improve the accuracy of CTR prediction. FPMC\cite{rendle2010factorizing} introduces a personalized transition matrix based on Markov chains, which captures both time information and long-term user preference information. YoutubeDNN\cite{covington2016deep} uses average pooling to encode user behavior sequences into a fixed-length vector to feed into MLP. DIN\cite{zhou2018deep} learns the user's historical behavior representation through the attention mechanism. DIEN\cite{zhou2019deep} further accomplishes efficient characterization of user behavior sequences by introducing auxiliary loss, and then uses AUGRU to capture the evolving trend of user interests. However, it should be pointed out that the time intervals of user behavior sequences are not evenly spaced, so RNN-based techniques are not perfectly suitable for this problem. SLi-Rec\cite{yu2019adaptive} improves the structure of LSTM and introduces time difference between the adjacent items to model unequally-spaced behavior, which significantly improves the performance.

\subsection{Session-based Deep Models}
In CTR prediction tasks, there are not many session-based deep methods. GRU4REC\cite{hidasi2015session} uses RNN for session-based recommendation for the first time, and the user's click sequence is compressed by embedding for the purpose of forming a continuous low-dimensional vector input to GRU. After that, neural attentive recommendation machine (NARM)\cite{li2017neural} was proposed to model the user's sequential behaviors and capture the user's main intent in the current session. DSIN\cite{feng2019deep} splits the user behavior sequence into several sessions, and innovatively uses the ``transformer''\cite{vaswani2017attention} to model the internal isomorphism inside a session and heterogeneity between sessions, and then captures the interaction between sessions through Bi-LSTM\cite{huang2015bidirectional}. 

\section{Knowledge Aware Adaptive Session Multi-Topic Network}
In this section, we elaborate the proposed algorithm. As shown in Figure \ref{fig:pathdemo2}, the KAST architecture consists of 4 modules:



\begin{itemize}
  \item Adaptive Session Segmentation module. It can optimize session division from the whole sequence automatically, where the intents in each session are maintained similar respectively. 
  \item Embedding module. It transforms sparse user-related and item-related features into fixed-length dense vectors. 
  \item Behavior Sequence Representation module. It models topic representation from user behavior sequence, and includes 4 layers from the bottom to up: 
  (1).Topic Distillation Layer distills a vector representation within each session; 
  (2).Topic Transfer Layer learns semantic information from the session-level sequence; 
  (3).KSE module extracts structural information from the user-item interaction graph in the form of auxiliary loss, which is incorporated with other information to improve the embedding accuracy, and guarantees that similar items are closer to each other in latent space; 
  (4).Topic Evolving Capture Layer applies the attention unit to weigh the interest of users in each session by evaluating correlation degree with the target item.
  \item MLP. Its inputs are the outputs of Activation Unit layer and embedding representation of users, items and relations.  
\end{itemize} 

\subsection{Adaptive Session Segmentation Module}


While most behaviors within the same session are strongly related to each other, as shown in Figure \ref{fig:pathdemo1}, some irrelevant user behaviors on the borders of adjacent sessions may also be included in this session, which deviates from the major intention of this session. In order to mitigate the noise effect, the ASS module is proposed to adaptively adjust the borders between adjacent sessions.  

At the beginning of training, the initial integrated sequences are segmented simply based on the time gap \cite{feng2019deep}. In this process, each user-touched item sequence $S$ is segmented into sessions $Q=\{Q_{1},Q_{2},...,Q_{n}\}$, where $n$ is the number of segmented sessions. Moreover, the $i$-th session $Q_i = \{b_1^i,b_2^i,...,b_{T_{i}}^i\}$, where $T_i$ is the number of user behaviors in $Q_i$ and $b_k^i$ is user's $k$-th item in $Q_i$.


The structure of ASS module is illustrated in Algorithm \ref{alg:algorithm}. The inputs are sessions $Q$. At the beginning of each iteration, $Q_{i}$ and $Q_{i+1}$ are fed to the ASS unit, and this unit will output the $Q_{i}$'s final result $Q_{i}^{'}$ and the $Q_{i+1}$' intermediate result $\widetilde{Q}_{i+1}$. Next, the intermediate result $\widetilde{Q}_{i+1}$ and the original $Q_{i+2}$ will be used as the input for the next ASS unit. The adaptive operation will be repeated until all sessions are processed.

The internal process flow and algorithm of the ASS module can be found in Algorithm \ref{alg:algorithm} respectively, and the details are discussed below. 

(1) The last $K$ behaviors of $Q_i$ and the first $K$ behaviors of $Q_{i+1}$ are extracted, where $K$ denotes the division depth. K would be adjusted according to different datasets, and the performance is not very sensitive to selection of K value. In the circumstance where the number of behaviors in a session is less than K, all behaviors in the session will be calculated. Here we denote $s_k^{i+1}$ as the $k$-th item one-hot representation of the first $K$ behaviors we extract in $Q_{i+1}$, and $e_k^{i}$ as the $k$-th item one-hot representation of the last $K$ behaviors we extract from $Q_i$, where $e_k^{i},s_k^{i+1} \in R^{M}$, and $k \in 1,2,...,K$. $s_k$, $e_k$ are transformed into low-dimensional dense vectors by embedding matrix as follows: 

\begin{small}
{\setlength\abovedisplayskip{0pt}
\setlength\belowdisplayskip{-10pt}
\begin{equation} 
v_k^{i+1} = {s_k^{i+1}}\cdot E_{iter} 
\end{equation}
\begin{equation} 
v_k^{i} = {e_k^{i}}\cdot E_{iter}
\end{equation}
}
\end{small}

\noindent where $E_{iter} \in R^{M \times d_{model}}$ is the item embedding matrix, $M$ is the size of total item set, $d_{model}$ is the embedding size, $iter$ is the training iteration rounds, and $v_k^i$, $v_k^{i+1} \in R^{d_{model}}$ is the embedding representation for $e_k^{i},s_k^{i+1}$. Furthermore, the average embedding vector $\bar S_i, \bar S_{i+1} \in R^{d_{model}}$ for the entire session $Q_i$ and $Q_{i+1}$ are utilized as embedding vector representation of these two sessions, and are calculated as follows:

\begin{small}
{\setlength\abovedisplayskip{1pt plus 3pt minus 10pt}
\setlength\belowdisplayskip{1pt plus 3pt minus 10pt}
\begin{equation}
\bar S_i = \frac{1}{N_i}\sum_{k=1}^{N_i}{{b_k^i}\cdot E_{iter}}
\end{equation}
\begin{equation}
\bar S_{i+1} = \frac{1}{N_{i+1}}\sum_{k=1}^{N_{i+1}}{{b_k^{i+1}} \cdot E_{iter}}
\end{equation}
}
\end{small}

\begin{small}
\begin{algorithm}[t]  
  \small
  \caption{Adaptive Session Segmentation}  
  \label{alg:algorithm}
  \KwIn{Set of sessions $Q_{i},Q_{i+1}$;  
      similarity threshold $\alpha$ and adaptive division depth $K$;  
      Similarity function $Sim(\cdot, \cdot) $;
      Item Embedding Matrix $E_{iter}$;}  
  \KwOut{Two sessions, $final\_{Q_i^{'}}$ and $temp\_\widetilde {Q}_{i+1}$}  
  $\bar S_i \leftarrow mean(Q_i)$; $\bar S_{i+1} \leftarrow mean(Q_{i+1})$\;  
  $e_k^i \leftarrow extract\_last\_kth\_element(Q_i, k,K)$\;
  $s_k^{i+1} \leftarrow extract\_first\_kth\_element(Q_{i+1}, k,K)$\;  
  $v_k^i \leftarrow s_k^i \cdot E_{iter}$; $v_k^{i+1} \leftarrow s_k^{i+1} \cdot E_{iter}$\;  
  \For{$k=1,...,K$}  
  {  
    ${\theta}_{i,i,k} = Sim(\bar S_i, v_k^i)$\;  
    ${\theta}_{i+1,i,k} = Sim(\bar S_{i+1}, v_k^i)$\;  
    ${\theta}_{i,i+1,k} = Sim(\bar S_i, v_k^{i+1})$\;  
    ${\theta}_{i+1,i+1,k} = Sim(\bar S_{i+1}, v_k^{i+1})$\; 
  \If{${\theta}_{i,i,k} < \alpha$ and ${\theta}_{i+1,i,k} > {\theta}_{i,i,k}$}
  {  
    insert $e_k^i$ to the head of $Q_{i+1}$, remove $e_k^i$  from $Q_{i}$\;  
  }  
  \If{${\theta}_{i,i+1,k} < \alpha$ and ${\theta}_{i+1,i+1,k} > {\theta}_{i,i+1,k}$}
  {  
    insert $s_k^{i+1}$ to the tail of $Q_i$, remove $s_k^{i+1}$  from  $Q_{i+1}$\;  
  }     
  }  
  return $final\_{Q_i^{'}}$, $temp\_\widetilde {Q}_{i+1}$\
\end{algorithm}
\end{small}

\noindent where $N_i$, $N_{i+1}$ are the item number for the entire session $Q_i, Q_{i+1}$, respectively, and $b_k^i, b_k^{i+1}$ are the $k$-th item of session $Q_i$ and $Q_{i+1}$.

(2) These two adjacent sessions are adjusted by calculating the forward and backward similarity simultaneously. Specifically, forward similarity determines whether to modify $Q_{i+1}$ given $Q_i$, and backward similarity determines whether to modify $Q_i$ given $Q_{i+1}$. There are many options for the similarity function, such as Euclidean distance and Cosine, etc. $Q_i$ can be determined by the current unit, but $Q_{i+1}$ still needs to be further processed in the next unit. The process is as follows:

\begin{small}
{\setlength\abovedisplayskip{1pt plus 3pt minus 5pt}
\setlength\belowdisplayskip{1pt plus 3pt minus 15pt}
\begin{equation}
{\theta}_{i,i,k} = Sim(\bar S_i, v_k^i), \qquad\qquad\qquad k \in 1,...,K
\end{equation}
\begin{equation}
{\theta}_{i+1,i,k} = Sim(\bar S_{i+1}, v_k^i), \qquad\qquad k \in 1,...,K
\end{equation}
\begin{equation}
{\theta}_{i,i+1,k} = Sim(\bar S_i, v_k^{i+1}), \qquad\qquad k \in 1,...,K
\end{equation}
\begin{equation}
{\theta}_{i+1,i+1,k} = Sim(\bar S_{i+1}, v_k^{i+1}), \qquad k \in 1,...,K
\end{equation}
}
\end{small}

\noindent Here $Sim(\cdot,\cdot)$ is the function that  calculates the similarity coefficient ${\theta}_{i,j,k}$ between two vectors, where $i$ denotes session $Q_i$, and $j,k$ denotes the vector of $Q_j$'s $k$-th behavior embedding $v_k^j$. For backward-unit, the similarity ${\theta}_{i,i,k}$ is calculated using the average embedding vector $\bar S_i$ and $v_k^i$. If ${\theta}_{i,i,k} < \alpha$, which means the i-th item is not correlative enough to the current session, we will move on to the next item, and the similarity ${\theta}_{i+1,i,k}$ would be calculated using $v_k^i$ and $\bar S_{i+1}$. If ${\theta}_{i+1,i,k} > {\theta}_{i,i,k}$, $e_k^i$ would be removed from $Q_i$ and inserted to the head of $Q_{i+1}$.
Similarly, for forward-unit, if ${\theta}_{i,i+1,k} < \alpha$, ${\theta}_{i+1,i+1,k}$ would be calculated, and if ${\theta}_{i+1,i+1,k} > {\theta}_{i,i+1,k}$, $s_k^{i+1}$ would be removed from $Q_{i+1}$ and inserted to the tail of $Q_i$. Here $\alpha$ denotes similarity threshold. The overall flow can be found in Algorithm \ref{alg:algorithm}. 


As is shown in Algorithm \ref{alg:algorithm}, the embedding matrix is fundamental for session division optimization in ASS module, and it is continuously updated as the algorithm iterates. To learn the latent representation which can fit with the relations between items from user behaviors, various information should be incorporated together. The KSE module, which provides the structural information, is integrated into the framework, and we will introduce it later. It should be noted that ASS module does not optimize those non-adjacent sessions, because we don't want ASS module to disrupt the order between sessions.

In order to obtain a better session segmentation performance, we must first warm up the model, and then fine-tune the sessions after the embedding vectors is relatively stable. The computation complexity of this module is $O(KN)$, and $N$ is the number of samples.

\subsection{Topic Distillation Layer}
After the session division for the whole sequence is optimized, the pooling operator is utilized to capture the major topic vector within each session. As depicted in Figure \ref{fig:pathdemo2}, each session is connected to a corresponding average pooling operator, which aggregates the information:

\begin{small}
{\setlength\abovedisplayskip{1pt plus 3pt minus 5pt}
\setlength\belowdisplayskip{1pt plus 3pt minus 15pt}
\begin{equation}
{topic}_i = Average\_Pooling(Q_{i}^{'})
\end{equation}
}
\end{small}

By applying the Average-Pooling operators, the model effectively distills the session topics. 

\subsection{Topic Transfer Layer}

We concatenate each topic representation into a session-level sequence, and the interest trends are much clearer from the aspect of the coarser granularity. Because GRU overcomes the vanishing gradients problem of RNN, and is more efficient than LSTM, it is applied here to model the evolving dependency across sessions. The formulations of GRU are listed as follows: 


\begin{small}
{\setlength\abovedisplayskip{1pt plus 3pt minus 5pt}
\setlength\belowdisplayskip{1pt plus 3pt minus 15pt}
\begin{equation}
h_i = \mathrm{GRU}(h_{i-1},topic_i;\Theta)
\end{equation}
}
\end{small}

\noindent where $\mathrm{GRU}(\cdot)$ is the GRU unit, $h_i$ the $i$-th hidden states, $topic_i$ the input of GRU, and $\Theta$ refers to all parameters related to GRU. Through this layer, user interest transferring trends that are incorporated with semantic information can be captured.


\subsection{Topic Evolution Capture Layer}
It is a fact that the correlation between a target item and several historical user interests can be distinct, and this is the reason why attention mechanism is employed. This mechanism implements soft alignment between sources and targets, and proves to be effective in weight allocation problems. Therefore, the weights between the target item and session interests can be learned, after which the weights are multiplied by the different interest representation to compute the final weighted interest representation. The adaptive representation of session interests w.r.t. the target item is calculated as follows: 

\begin{small}
\begin{equation} 
a_i = \frac{exp(IW^{att}h_i)}{\sum_{SN}exp(IW^{att}h_i)}
\end{equation}
\begin{equation} 
U = \sum_{SN}a_{i}h_{i}
\end{equation}
\end{small}

where $I\in R^{n_I}$ is the target item's embedding vector, $W^{att}\in R^{n_I \times n_H}$, $n_I$ the input size, $n_H$ is the hidden size, $h_i$ is the output of GRU unit, and $SN$ is the session number. Next, each $h_i$ will be multiplied with its corresponding weight $a_i$ to compute the output $U$, and then it will be concatenated and fed to the MLP layer.

\subsection{Knowledge aware Structural-information Extraction Module}
It is obvious that the embedding vector is fundamental for ASS module. To enhance the embedding characterization, KSE module is proposed, which introduced the structural information of the knowledge, characterize embedding vector in different domains. KSE module learns a low-dimensional vector between each entity and relationship, and extracts the structural and semantic information of the original graph in the vector. In this way, the mutual relations between entities are well represented by the vectors learned by KSE module. A typical knowledge graph ($KG$) consists of millions of entity-relation triples $(h, r, t)$, in which $h$, $r$ and $t$ represent head entity, relation, and tail entity of a triple, respectively. It is worth noting that in existing methods\cite{huang2018improving,wang2018dkn} $KG$ embedding is pre-trained and the embedding results are fed into final CTR prediction model. However, In KAST, KSE embedding and the prediction model are integrated into one framework and trained in an end-to-end manner. as shown in Figure \ref{fig:pathdemo2}.


For the KSE module, we use model TransE\cite{bordes2013translating} to minimize the distance-based score function: $f_r(h,t) = \parallel h+r-t \parallel_2^2$, where $h$, $r$, $t$ form a triplet. Each clicked sample is labelled positive. We randomly sample 5 triplets from the same batch as negative triplets, and use $\Phi$ and ${\Phi}^{'}$ to denote a positive triplet and the corresponding negative one, respectively. Then we define the following margin-based objective as the auxiliary loss : 
\begin{small}
\begin{equation}
L_{KSE} = \sum_{\Phi\in \triangle}\sum_{{\Phi}^{'} \in {\triangle}^{'}}[\xi + f_r({\Phi}^{'}) - f_r(\Phi)]_+
\end{equation}
\end{small}
where $[x]_+=max(x, 0)$, $\xi$ is the margin separating positive and negative triplets, and $\triangle = \{(h_j, r_j, t_j)\,|\,positive\}$, ${\triangle}^{'} = \{(h_j, r_j, t_j)\,|\,negative\}$ respectively. We use $L_{KSE}$ as the loss function of the KSE module. The global loss for KAST is 
\begin{small}
\begin{equation}
L = L_{pCTR} + \gamma \ast L_{KSE}
\end{equation}
\end{small}

where $\gamma$ is the hyper-parameter which balances the main CTR prediction loss $L_{pCTR}$ and the auxiliary loss $L_{KSE}$. With the help of the KSE module, we are able to obtain rich structural information of item embedding from a variety of domains. Specifically, in addition to TransE, we also exploit the idea from TransH~\cite{wang2014knowledge} and TransD~\cite{ji2015knowledge}.TransH introduces a relation hyperplane matrix to optimize the relationship of triplets, and TransD uses the product of two projection vectors of a triplet to map entity embeddings to the corresponding relation space, which is known as the $KG$ Embedding methods.



\section{Experiments}
\subsection{Experimental setup}

\subsubsection{Datasets} We select two datasets for the following experiments, and they are both typical scenarios of CTR prediction models. (1) Alibaba Dataset\footnote{\url{https://tianchi.aliyun.com/dataset/dataDetail?dataId=56}}: This is an advertising dataset released by Alimama, an online advertising platform of Alibaba. Logs from 2017-05-06 to 2017-05-12 are for training and logs from 2017-05-13 are for testing.
(2) MovieLens Dataset\footnote{\url{https://grouplens.org/datasets/movielens/}}: The MovieLens datasets contains rating on recommended movies by users, as well as movie metadata information and user attribute information. Samples rated greater than 3 are regarded as positive, and the rest are negative samples. We take the first 11 months as the training set and the last month as the test set. The statistical information is listed in Table \ref{tab2}.

\begin{table}[htbp]
\setlength{\abovecaptionskip}{1pt} 
\setlength{\belowcaptionskip}{-10pt}
\centering
\small
\begin{tabular}{ccccc}
               \hline
                 Dataset & $\sharp$Users & $\sharp$Items & $\sharp$Relations & $\sharp$Instances\\
               \hline
                 Alibaba Ads & 63706 & 281859 & 3 & 1483560\\
               \hline
                 MovieLens & 12517 & 8509 & 19 & 1803158\\
               \hline
             \end{tabular}
             \caption{The statistics of datasets} 
             \label{tab2}    
\end{table}

\subsubsection{Methods to compare} We compare the proposed method with the following competitive ones: LR, FM, Wide\&Deep, DCN, YoutubeDNN, AFM, DeepFM, AutoInt, PNN, DIN, DIEN, DSIN, and these model are introduced in Sections 1 and 2. Besides, we designed a GRU model based on the Embedding\&MLP, where the user behavior sequence is processed by GRU, named GRU-net. And similarly, LSTM-net, where LSTM module replaces GRU module.

\subsubsection{Evaluation Metrics} Since we formulate the CTR prediction task as a binary classification problem, $AUC$ and $loss$ are used as the evaluation metrics. $AUC$ measures the model's ability to rank a positive instance higher than a negative one.
$loss$ is the the averaged cross entropy for all the sample, and it does not include any regularization term. 

\subsubsection{Parameter Settings} 
In order to ensure consistency, we use the same features of both datasets, and use sliding window to extract user historical behavior features. On these two datasets, we set (1) embedding size of all models to 24. (2) Fully connected layer settings, DCN, DeepFM is [128,128], AutoInt is [256, 256], GRU-net, LSTM-net, Wide\&Deep, YoutubeDNN, PNN, DIN, DIEN, DSIN, KAST is [200, 80]. (3) Optimizer for all models is Adam~\cite{kingma2014adam}. (4) Use Batch normalization~\cite{ioffe2015batch} for all models. (5) In order to ensure fairness, we also add user behaviors by sum-pooling for Wide\&Deep, DCN, YoutubeDNN, AFM, DeepFM, and AutoInt for comparison, and all model’s user sequences take the most recent 100 actions. (6) Session numbers for DSIN are set as 8.

\begin{table}

\small

\label{Tab03}
\begin{center}
\begin{tabular}{ccccc}

\toprule
Model&Alibaba Dataset &MovieLens Dataset \\

\cmidrule(r){2-3} 
\cmidrule(r){4-5}

&Original \quad + behavior     &Original \quad + behavior             \\

\midrule

LR  &0.6054 \quad - &0.7260 \quad -        \\

FM  &0.6081 \quad - &0.7669  \quad -       \\

\midrule

Wide\&Deep  &0.6087\quad0.6195 &0.6900\quad0.7953            \\

DCN  &0.6100\quad0.6201 &0.7760\quad0.7613            \\

YoutubeDNN  &0.6085\quad0.6282 &0.6525\quad0.7910            \\

AFM &0.6104\quad0.6144  &0.7396\quad0.7423            \\

DeepFM  &0.6188\quad0.6213  &0.8010\quad0.8058       \\

AutoInt  &0.6137\quad0.6214  &0.7609\quad0.7947    \\

PNN  &- \quad 0.6260        &- \quad 0.7444           \\

\midrule

GRU-net             &0.6243                               &0.7943                       \\

LSTM-net            &0.6228                     &0.7905                         \\

DIN               &0.6251                               &0.7453                       \\

DIEN              &0.6257                               &0.7750                       \\

DSIN              &0.6261                             &0.8169                       \\

\midrule

KAST              &$\mathbf{0.6330}$                    &$\mathbf{0.8219}$          \\

\bottomrule

\end{tabular}
\caption{Model performance (AUC) on public datasets}
\label{tab3}
\end{center} 
\end{table}

\begin{figure}[t]
\setlength{\abovecaptionskip}{0pt} 
\setlength{\belowcaptionskip}{-10pt}  
\centering
\subfigure[Alibaba Dataset]{
\includegraphics[width=3.5cm]{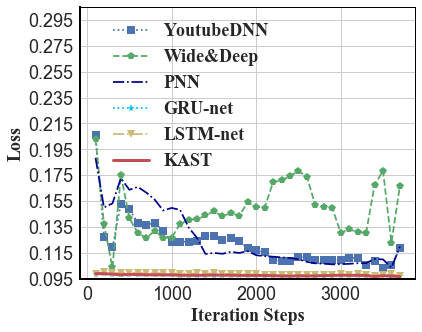}}
\subfigure[Movielens Dataset]{
\includegraphics[width=3.5cm]{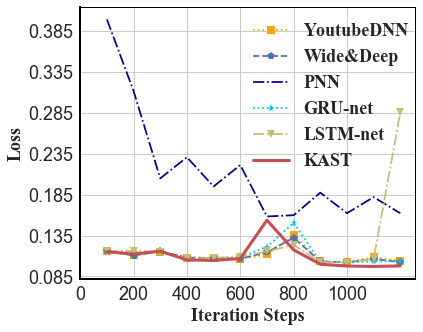}}
\caption{Logloss comparison of different dataset}
\label{fig:pathdemo6}
\end{figure}

\subsection{Performance Comparisons with Baselines}

\begin{figure}[t]
\centering
\subfigure[$\mathrm{Test\;AUC^1}$]{
\includegraphics[width=3.5cm]{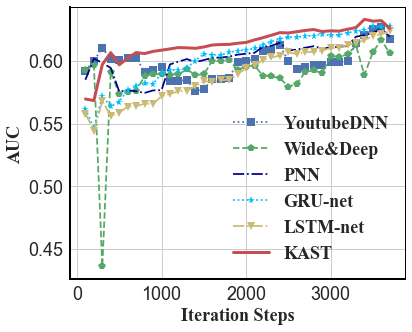}}
\subfigure[$\mathrm{Test\;AUC^2}$]{
\includegraphics[width=3.5cm]{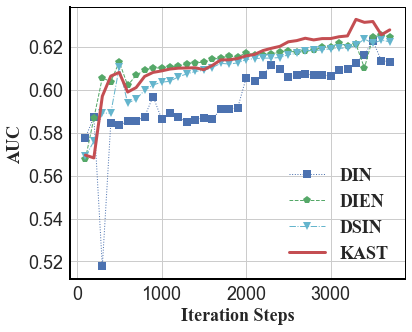}}
\centering
\subfigure[$\mathrm{Test\;AUC^1}$]{
\includegraphics[width=3.5cm]{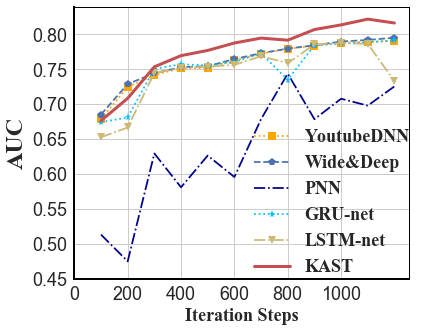}}
\subfigure[$\mathrm{Test\;AUC^2}$]{
\includegraphics[width=3.5cm]{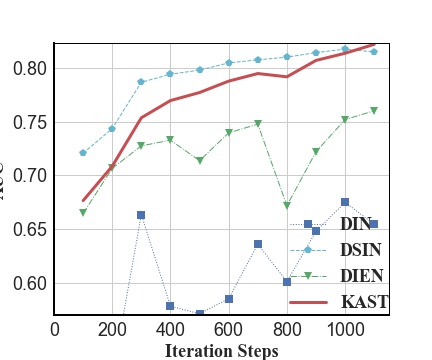}}
\caption{AUC comparison between KAST and compared methods on Alibaba(a-b), MovieLens(c-d) datasets}
\label{fig:pathdemo7}
\end{figure}

We divide models into three groups: non-sequence-based models, sequence-based models, and the latest models (DIN, DIEN, DSIN). Additionally, LR and FM are also selected as the classical models in this evaluation. The performances of these methods are illustrated in Table \ref{tab3}. Not surprisingly, the performance of LR and FM is the worst among all the methods. For Wide\&Deep, DCN, YoutubeDNN, AFM, DeepFM, AutoInt, we implement two versions: non-user-behavior version and user-behavior-enhanced version. In the later version, a user behavior extraction module is incorporated into the original network by embedding each item of user behavior sequence into dense vectors and pooling them into one vector. It is clearly shown that the performance of the enhanced version is augmented compared with the original one on both datasets. Hence, It is also proven that user behavior information is valuable to understand the user intention. Compare to the user-behavior-enhanced version of above-mentioned methods, the sequence-based methods (GRU-net and LSTM-net) are more powerful. The reason is that learning the migration process of user interests can obtain much finer-grained knowledge for predicting user intent than just aggregating the behavior. However, as elaborated in Section 3, when we learn the interest evolution on the finest-grained items, the noise among the user behavior sequence may disturb the intent understanding. In contrast, by adaptively segmenting the sequence during each iteration, the proposed method is able to learn the unnoised or less-noised interest evolution in the session granularity. The loss and AUC curves in Figures \ref{fig:pathdemo6} (a-b) and Figures \ref{fig:pathdemo7} (a,c) substantiate the effectiveness of session-level interest evolution learning on both datasets. 
Compared with the above-mentioned methods, DIN, DIEN, and DSIN went much deeper to mine user behavior sequences. As depicted in Figures \ref{fig:pathdemo7} (b) and (d), we compare KAST with them on two datasets. KAST yields the best performance, and the order is KAST\textgreater DSIN \textgreater DIEN \textgreater DIN. DIN does not use the sequential information of user behaviors, and the attention mechanism is only applied in item level. DIEN improves DIN by incorporating GRU with attention mechanism. However, the user interest representation is still based on item-wise level, and some noise is unavoidably included into the representation. DSIN learns the user interest evolution on session level. As shown in Table \ref{tab3} and Figures \ref{fig:pathdemo7} (b) and (d), the AUC curve of DSIN are indeed better than DIEN and DIN. However, as sessions are divided just by time interval, some items may not be arranged into the most appropriate session. KAST can adaptively refine segmentation of the sessions by similarity between the embedding vectors of items. Moreover, the user-item-interaction-based knowledge graph further enhances the effect of embedding learning. Hence, by simultaneously extracting session level user interest evolution and behavior structure information, the proposed method can accomplish better performance than other methods.


\subsection{Evaluation for Key Modules in KAST}
\subsubsection{Effect of ASS}  
As shown in Table \ref{tab4}, after the ASS module is added, results on both datasets are improved. In addition, it is proved that ASS can enhance session-wise quality for the user interest representation, After ASS module fine tunes the interval of session division, the items will be segmented to the most correlative topic, guaranteeing that the interest evolution learning of Topic Transfer Layer can be less affected by noises. When volume of items is large and the user-item behaviors are various, ASS module would be crucial.

\begin{table}[t]
\centering
\small
\begin{tabular}{ccc}
               \hline
                 Model & Alibaba & Movielens\\
               \hline
                 KAST without ASS and KSE & 0.6303 & 0.8185\\

                 KAST with ASS & 0.6313 & 0.8199\\

                 KAST with ASS and TransE & 0.6330 & 0.8204\\

                 KAST with ASS and TransH & 0.6311 & 0.8219\\

                 KAST with ASS and TransD & 0.6315 & 0.8177\\
               \hline
             \end{tabular}
             \caption{Comparison of model performance with ASS module and different KSE methods.}
             \label{tab4}  
\end{table}

\subsubsection{Effect of KSE}
It is shown in Table \ref{tab4} that KSE module can further improve performance. After KSE module is introduced in a form of an auxiliary loss. 
We applied three $KG$ embedding methods, TransE, TransH, and TransD, and compared the performances with and without these KG embedding methods, proving that KSE can further improve the performance. This result reflects the importance of high-quality embedding when performing session refinement and subsequent interest evolution learning. On the Alibaba datasets, TransE yields the best performance, as a result of low embedding size and just 3 relationships. TransE has a relatively simple structure and few parameters, which reduces the risk of overfitting. However, on the MovieLens datasets, where the number of relations is increased to 19, TransH performs the best. A possible explanation is that TransH has an additional matrix, which is able to map the vectors into a proper hyperplane. In both datasets, TransD did not perform the best, due to its large amount of parameters and thus likely to overfit. 




\subsection{Analysis of Hyper-parameters}
\subsubsection{Effect of session number SN} 
\begin{figure}[t]
  \setlength{\abovecaptionskip}{0pt} 
  \setlength{\belowcaptionskip}{-10pt}
  \centering
  \subfigure[Movielens Dataset]{\includegraphics[width=3.89cm]{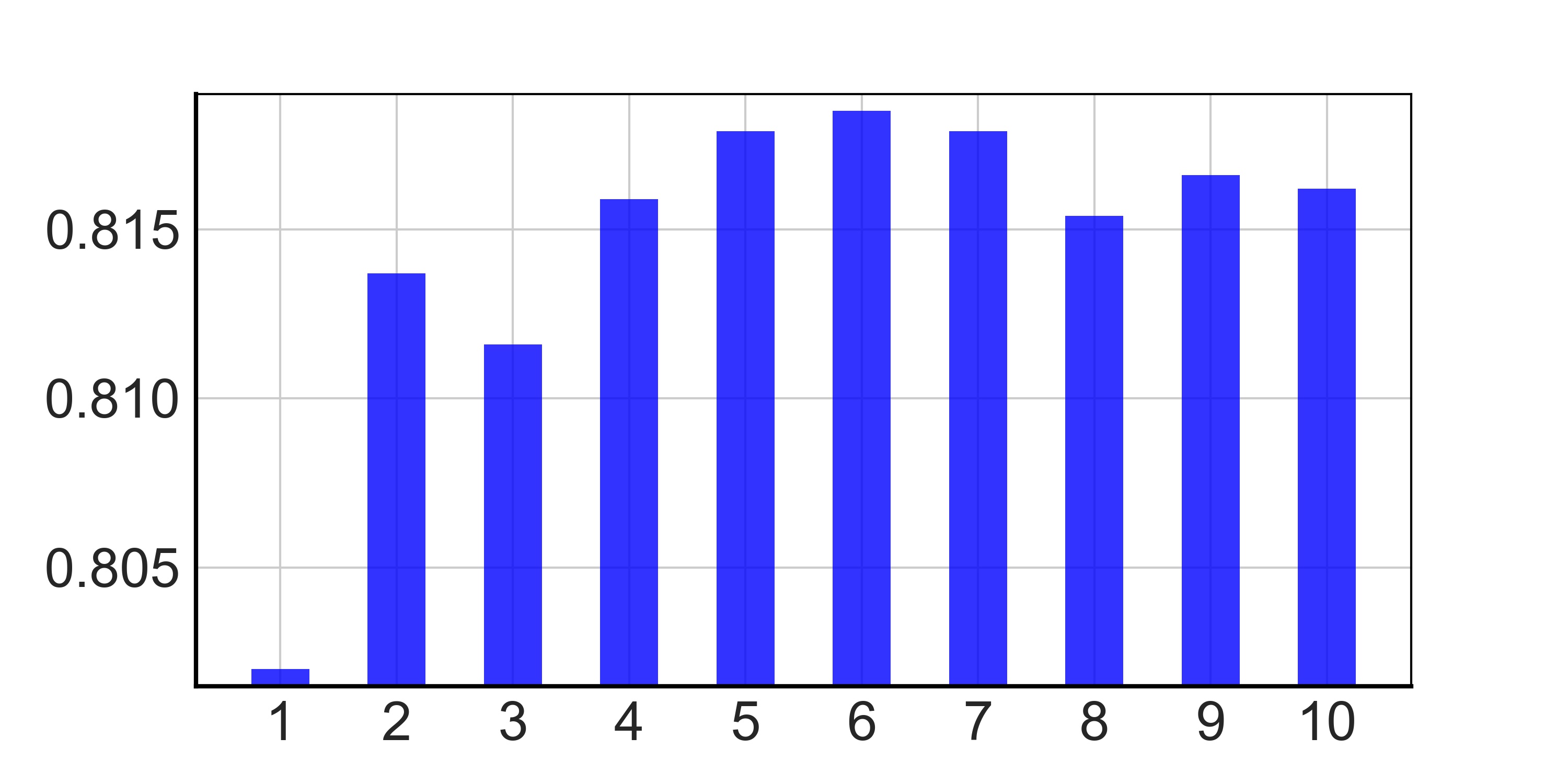}}
  \subfigure[Alibaba Dataset]{\includegraphics[width=3.89cm]{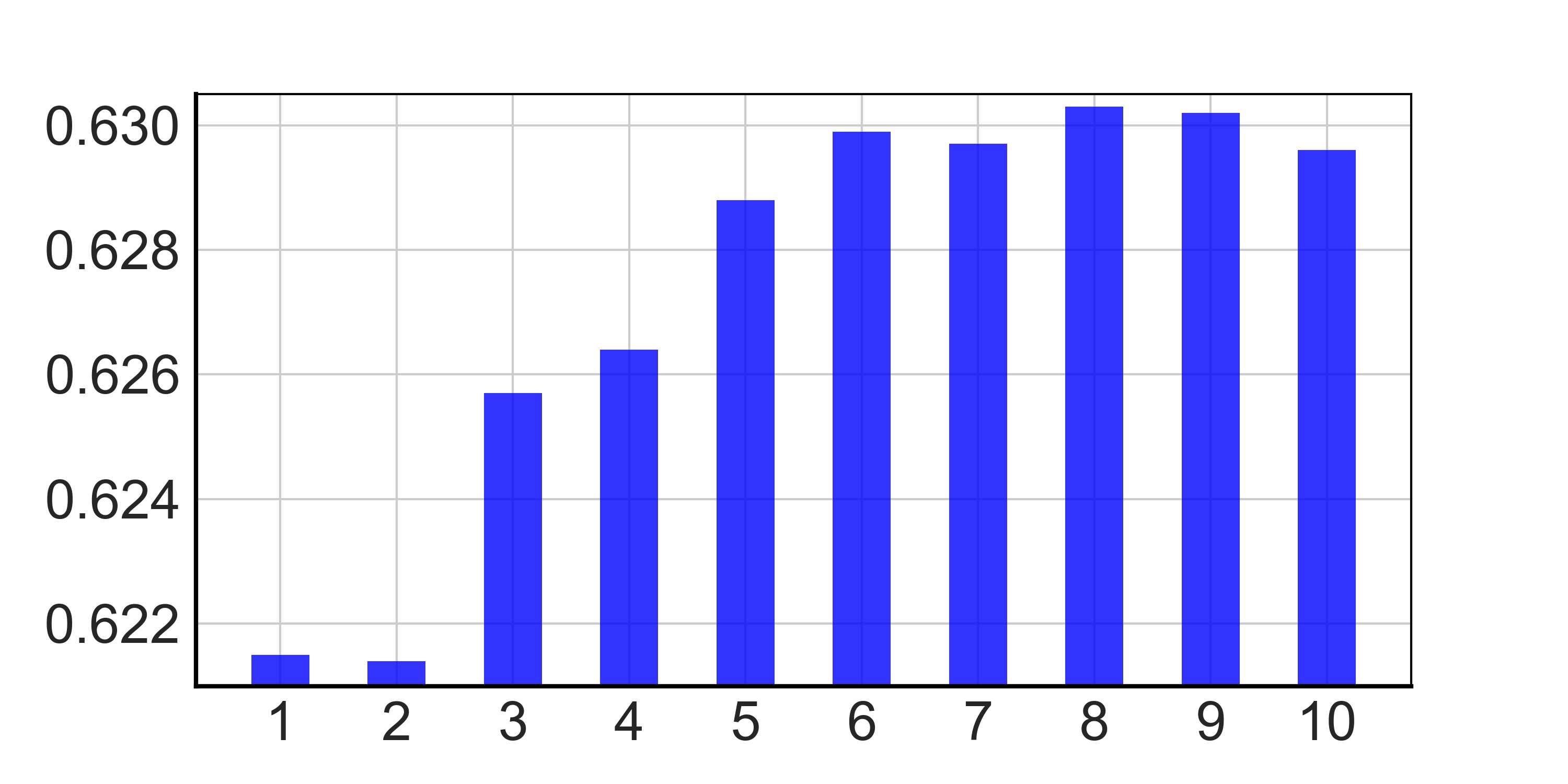}}
  \caption{Effect of session number $SN$.}
  \label{fig:pathdemo9}
\end{figure}

It manifests in Figure \ref{fig:pathdemo9} that the proposed method is not very sensitive to the selection of session number, and the value can be chosen from a relatively wide range. However, overly few sessions will lead to incomplete coverage of user interest points. On the other hand, if the sessions are too many, we should adjust the maximum length of each session, otherwise the sessions at the tail will be empty and then filled with padding.  When there are too many empty sessions, noise will be introduced, resulting in effectiveness decay. Generally, according to the richness of the user's behaviors in the datasets, 5 to 10 sessions are taken as optimal. On the Alibaba datasets, we take the number as 8. Compared to taking only one session, AUC of the basic KAST is increased by 1.42\%. In the MovieLens datasets, we take the number as 6, compared to taking only one session, AUC of basic KAST increased by 2.06\%.  

\begin{figure}[t]
  \setlength{\abovecaptionskip}{0pt} 
  \setlength{\belowcaptionskip}{-10pt}
  \centering
  \subfigure[Movielens Dataset]{\includegraphics[width=3.89cm]{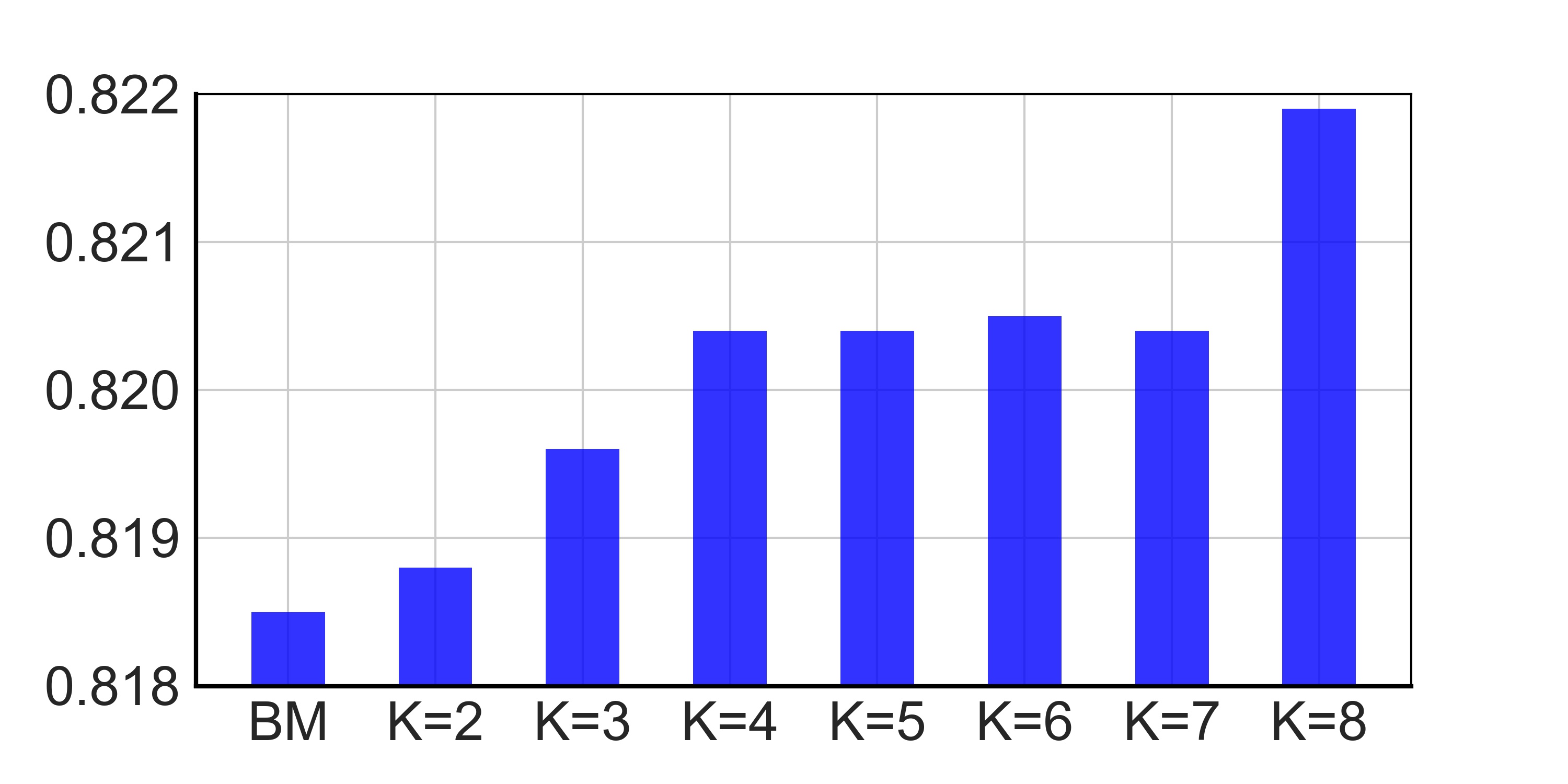}}
  \subfigure[Alibaba Dataset]{\includegraphics[width=3.89cm]{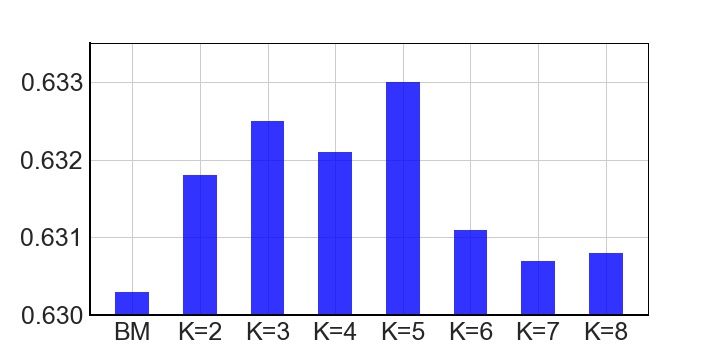}}
  \caption{Effect of $K$ value, where BM denotes KAST without ASS and KSE.}
  \label{fig:pathdemo10}
\end{figure}

\subsubsection{Effect of extraction element number K}  The results are summarized in Figure \ref{fig:pathdemo10}. It is clear that the item number a session has an impact on the model performance. The reasons may be as follows. $(1)$ If the user behaviors of the dataset are scattered, $K$ should be set large, so more misclassified items can be corrected to the most correlated sessions. $(2)$ If the user interest in a session is more concentrated, $K$ should be set smaller, because as user behaviors are to be concentrated in such period, the proportion of misclassified behaviors will be relatively small. On the Alibaba datesets, user's behaviors are concentrated, of which the neighboring ones are mostly of the same interest. However, on the MovieLens datasets, users may click on various types of movies in a short period of time. To cope with the latter occasion, we should set a larger value of $K$ to fully explore more casual behaviors in the session, and vice verse. Hence, the performance of KAST peaks when $K=5$ for for Alibaba dataset, and $K=8$ for for MovieLens dataset. 

\section{Conclusions}
In this paper, we propose a CTR prediction framework, namely KAST, that captures user interest topics by adaptively dividing user sequence into sessions. Specifically, after dividing user sessions, KAST captures session-level topics of user interest through the pooling layer, and then uses GRU to capture user topic transfer trends. KAST well alleviates the problem that the user's behavior sequence is not equally spaced. Furthermore, it transforms the behavior-related-items to the topic of interest, and uses these topics to compute the interaction with the target item, which enables KAST to capture the most perceptual macro topics of interest. In order to improve the quality of session division and representation, a knowledge aware module is proposed to extract the structural information from users behaviors, and this margin-based loss is merged into the major loss function. Experiments are conducted and demonstrate that KAST achieves superior performance on public datasets. For future work, we will design new modules to incorporate more information about user’s behavior sequence and user-item graph, thus to further improve the ranking performance.

\bibliographystyle{aaai21}
\bibliography{author-ref}

\end{document}